\documentclass[12pt,a4paper]{article}
\usepackage{jheppub}
 \usepackage{fixltx2e}
\usepackage{float}
\usepackage{amssymb,bm,latexsym,amsmath}
\usepackage{epstopdf}
\usepackage{caption}

 \usepackage{graphicx}
  \usepackage{epstopdf}
\usepackage[utf8]{inputenc}

\newcommand{\bej}[1]{ \begin{equation}\label{#1} }
\newcommand{\eej}{\end{equation}}
\newcommand{\beaj}[1]{\begin{eqnarray}\label{#1} }
\newcommand{\eeaj}{\end{eqnarray}}
\newcommand{\eq}[1]{(\ref{#1})}
\def\ZZZ{{\hskip-3pt\hbox{ Z\kern-1.6mm Z}}}
\def\zzz{{\hskip-3pt\hbox{ z\kern-1mm z}}}

\newcommand{\bd}{\bar{\rm D}}

\newcommand{\N}{\frac{m_{2}}{k_{2}}-\frac{m_{1}}{k_{1}}}

\newcommand{\be}{\begin{equation}}
\newcommand{\ee}{\end{equation}}
\newcommand{\ben}{\begin{eqnarray}\displaystyle}
\newcommand{\een}{\end{eqnarray}}

\def\one{{\hbox{ 1\kern-.8mm l}}}
\def\zero{{\hbox{ 0\kern-1.5mm 0}}}

\def\be{\begin{equation}}       
\def\ee{\end{equation}}         

\def\bea{\begin{eqnarray}}      
\def\eea{\end{eqnarray}}
                                
\def\ba{\begin{array}}
\def\ea{\end{array}}
\def\bd{\begin{displaymath}}
\def\ed{\end{displaymath}}

\def\eq{\begin{equation}}
\def\eqe{\end{equation}}
\def\eqa{\begin{eqnarray}}
\def\eqae{\end{eqnarray}}
\def\ena{\end{eqnarray}}

\def\nn{\nonumber}

\def\unit{1 \hskip-.3em \raise2pt\hbox{$ \scriptstyle |$ } }
\def\g{\gamma}

\def\l{\lambda}

  \def\w{\omega}
\def\s{\sigma}                                   %     \varsigma
\def\t{\tau}

\def\del{\partial}              
\def\bd{\begin{displaymath}}
\def\ed{\end{displaymath}}

\def\6{\partial}
\def\N4{{\cal N}=4}

\def\bop#1{\setbox0=\hbox{$#1M$}\mkern1.5mu
        \vbox{\hrule height0pt depth.04\ht0
        \hbox{\vrule width.04\ht0 height.9\ht0 \kern.9\ht0
        \vrule width.04\ht0}\hrule height.04\ht0}\mkern1.5mu}
\def\>{\rangle} %right angle

\def\<{\langle} %left angle
\def\Dsl{D \hskip-.6em \raise1pt\hbox{$ / $ } }
\def\to{\rightarrow}

\def\del{\nabla}

\def\+{\oplus}

\def\as2{AdS_3\times S^3_1 \times S^3_2}

\title{Multi-spike strings in $AdS_3$ with mixed three-form fluxes}

\author{Aritra Banerjee* and}
\affiliation{*Department of Physics\\ IIT Kharagpur\\ Kharagpur 721302, India} 

\author{Abhishake Sadhukhan**}
\affiliation{ **Harish-Chandra Research Institute,\\
 Chhatnag Road, Jhusi, \\ Allahabad 211019, India}

\emailAdd{ aritra@phy.iitkgp.ernet.in, abhishakesadhukhan@hri.res.in}

\abstract{
The string sigma model in $AdS_3\times S^3$ supported by mixed three-form fluxes  has recently been proved to be integrable which led to a plethora of work in this background including proposals for S matrix and study of  semiclassical string profiles. Motivated by this, in this paper, we present a study of `spiky' strings in this background. We analyze the string profiles in detail and also find the dispersion relation between the charges in the `long' string limit after solving the equations of motion perturbatively upto the leading order in the Neveu-Schwarz flux $b$. We find that the dispersion for 2 spikes gets corrected by the term $-\frac{b^2}{2}\log S$. We also discuss the fate of the solution in the limit of pure NS-NS flux. }

\begin{document}
\maketitle
\section{Introduction}\label{intro}

The $AdS/CFT$ correspondence\cite{Maldacena:1997re,Witten:1998qj,Gubser:1998bc} which states that string theory in $AdS$ background in $d$ dimensions is equivalent to a conformal field theory in $d-1$ dimensions has opened up a fascinating possibility to solve the non trivial dual field theory in the planar limit. This duality also relates the
various observables of the CFT such as anomalous dimension of gauge theory operators to the dispersion relation of the dual string states. Proving the duality in general has been highly challenging since the string theory side contains an infinite tower of states. However, in the so called large charge limit the problem becomes more tractable and it was observed that the anomalous dimensions of certain operators in the dual Super Yang-Mills (SYM) side can be related directly to the dispersion relation between conserved charges for particular classical strings in the $AdS$ background in certain limits \cite{Gubser:2002tv,Berenstein:2002jq}. This together with the remarkable observation that the dilatation operator of the SYM theory at one loop can be written exactly as the hamiltonian of a integrable spin-chain\cite{Minahan:2002ve} has been a key concept in connecting integrability, strongly coupled systems and string theory in the context of classical solutions. Various of the spin-chain excitations have a dual description in terms of the  string states in $AdS$. Hence it is expected that the classical string solutions in various $AdS$ backgrounds will help in our understanding of the dual field theory.

The most studied form of the $AdS/CFT$ correspondence, i.e. the duality between $\mathcal{N}$=4 SYM and type IIB superstrings in $AdS_5\times S^5$ has prompted a lot of studies on the semiclassical rotating string solutions arising out of the $AdS$ string sigma model(see \cite{Plefka:2005bk} for an early review). The most famous example of this might be that of the folded spinning string or GKP string found in\cite{Gubser:2002tv}. The dispersion relation for such a string in the conformal gauge, moving in $AdS_3\subset AdS_5$ for large value of the spin $S$ is given by 
\be
E - S = \frac{\sqrt{\lambda}}{\pi} \log S,
\ee
 This classical description of the string is valid in  the limit $\sqrt{\lambda}=\frac{R^2}{\alpha'}>>1$. This string solution has been found to be dual to twist two operators of the form $\text{Tr}(\Phi \del_+^S \Phi)$, $\Phi$ being the general complex scalar in the SYM theory and $\del_+$ is the covariant light cone derivatives. This relation has been confirmed from the perturbative calculations in the SYM side for this operator which provides a perfect match with the cusp anomalous dimension $f(\lambda)$ of Wilson line\cite{Drukker:1999zq,Kruczenski:2002fb,Makeenko:2002qe}, the coefficient of the logarithmic divergence in $S$. In this connection, the single trace operators have become an important object to be studied in the correspondence. Obviously, the most interesting contender here would be the general form of the operator
\be
\text{Tr}(\del_+^{S_1}\Phi_1 \del_+^{S_2} \Phi_2...\del_+^{S_n} \Phi_n)
\ee
i.e. higher twist operators with arbitrary number of fields, where the derivative contribution dominates in the large spin limit. In \cite{Kruczenski:2004wg}, it was shown that the string solution dual to such a operator has $n$ cusps or `spikes' and the symmetric case with $S_1=S_2=...=S_n$ was discussed. From the worldsheet viewpoint, the spikes (or cusps) are defined as a discontinuity in the unit spacelike tangent vector on the string, which does not hamper the criterion of a smooth worldsheet. It was also shown that keeping in with the expectation, the dispersion relation of such a string reduces exactly to that of the GKP folded string when $n=2$. A lot of studies have followed this finding and generalized `spiky' string solutions in various $AdS$ backgrounds \cite{Kruczenski:2006pk,Kruczenski:2008bs,Ishizeki:2008tx,Jevicki:2009uz,Kruczenski:2010xs} as well as integrable deformations thereof \cite{Banerjee:2015nha} have been reported. The same dispersion relation obtained from an integrable SL(2) spin chain was reported in \cite{Freyhult:2009bx}. The spin chain description of such spiky strings were explored in detail in the series of works \cite{Dorey:2008vp,Dorey:2010iy,Dorey:2010id,Dorey:2010zz}.

The other much discussed form of gauge-gravity duality comes in the form of strings in $AdS_3\times S^3\times M^4$ and the two dimensional CFTs. The classical integrability of this theory with RR three form fluxes and $M^4=T^4 ~\text{or}~ S^3\times S^1$ has been explored from both sides of the duality \footnote{For a recent review of integrability in $AdS_3/CFT_2$ one can see \cite{Sfondrini:2014via} and references therein.} and various classical string solutions \cite{Maldacena:2000hw, Lee:2008sk, David:2008yk, David:2010yg, Chen:2008vc, Abbott:2012dd, Beccaria:2012kb, Rughoonauth:2012qd, Sundin:2013ypa, Abbott:2013mpa} have also been constructed. It has recently been found that string theory in $AdS_3\times S^3\times T^4$ supported by both NS-NS and RR fluxes is integrable \cite{Cagnazzo:2012se}. Moreover,  there have been proposals of S-matrix in $AdS_3\times S^3$ in presence of these `mixed' fluxes based on integrability of the sigma model \cite{Hoare:2013pma,Hoare:2013ida,
Bianchi:2014rfa,Hoare:2013lja,Babichenko:2014yaa,Borsato:2014hja}. The three form fluxes in this background are  parameterized by $b$ which is the coefficient of (NS-NS) flux. The coefficient of RR flux is then fixed from the supergravity equations $\hat{b}=\sqrt{1-b^2}~$. Depending on the values of these numbers, the string sigma model interpolates between that of the $SL(2,\mathbb{R})$ WZW model (pure NS-NS) and the one supported only by RR fluxes.
% The dual field theory for the pure RR case is the $\mathcal{N}$= (4,4) superconformal field theory associated with the D1-D5 system.
%However the intermediate regime $b\neq 0$ remains mysterious since there is no reason to assume an underlying spin-chain structure. 
Various classical string solutions have been constructed in this regime, including the giant magnons and its finite size corrections \cite{Hoare:2013lja,Banerjee:2014gga,Ahn:2014tua}, folded string \cite{David:2014qta} and pulsating string solutions \cite{Banerjee:2015bia}, among others \cite{Hernandez:2014eta,Hernandez:2015nba}. The  field theory dual to this background exists and it's the ${\cal N} = (4, 4)$ superconformal field theory associated with the D1-D5 system. Motivated by these developments, we take up the task of studying the spiky strings in this background. 

As already mentioned, the spinning folded string in $AdS_3\times S^3$ with mixed flux background was constructed in an earlier paper by solving the equations of motion of the classical folded spinning string solution in terms of a perturbative expansion in the NS-NS flux parameter $b$. The dispersion relation there was found using perturbative expansion in  small $b$ and then taking the large spin limit ($S\to \infty)$ and has the following form written  up to the $\mathcal{O}(b^2)$,
\be
E-S=\frac{\sqrt{\lambda}}{\pi} \log S-\frac{\sqrt{\lambda}}{2\pi} b^2 (\log S)^2+\mathcal{O}(b^2 \log S),
\ee
so that the corresponding cusp anomalous dimension receives corrections due to the inclusion of the NS-NS flux. The $(\log S)^2$ term here is not suppressed by $1/S$ as is the usual case in SYM theories and remains relevant in the large $S$ limit. In the present paper we concentrate on the spiky string solution in the mixed flux background by following the same approach by solving the string equations of motion and expanding the Noether charges for a small flux parameter $b$. We find that the dispersion relation for the `long' spiky string is not equivalent to that of the folded string in this background even for $n=2$. We will show that the corrections in this case also contains a $\mathcal{O}(b)$ term which vanishes when $n=2$, however we find no $(\log S)^2$ term in the leading order of large $S$ limit, which points to the fact that in a general setting with background fluxes the equivalence of these two kind of strings breaks down and may correspond to completely different objects in the dual gauge theory. We will briefly discuss the subtle factors that may be at play here.

This short paper is organized as follows. In section \ref{spinning}, we will discuss the spiky string solutions in $AdS_3$ with mixed fluxes and find the relevant dispersion relation between the conserved charges. We will also describe the effective dynamics of such string analytically and pictorially. In section \ref{pure}, we describe the case of the pure NS-NS flux, i.e. $b=1$, where we would find an unique exact dispersion relation between the charges. In section \ref{conclusions}, we conclude our discussion and present further outlook.

 \section{Spinning spiky string with mixed fluxes}\label{spinning}
 
Let us first start with the $AdS_3$ background having a metric
\be
ds_{AdS_3}^2=-\cosh^2{\rho} dt^2+d\rho^2+\sinh^2{\rho} d\phi^2
\ee
and the NS-NS  and RR fields as following
\be\label{bans}
B_{t\phi}=b\sinh^2\rho;~~~C_{t\phi}= \sqrt{1-b^2}\sinh^2\rho.
\ee
where $0\leq b \leq 1$. Since we would be talking about a F-string solution, the RR flux will not be of importance to us. 

Note that there is a gauge freedom in the choice of the two-form B-field
for the Wess-Zumino term since the supergravity equations of motion depend on the three-form field strength, $H^{(3)}=d B$. We could have added a constant term $\xi$ to $B_{t\phi}$ in (\ref{bans}) but it will not change the string equations of motion, only change the conserved charges by an extra constant term proportional to $b \xi$. For open string solutions, the constant may affect the boundary conditions imposed on the string solutions. However, the string solutions we are interested in here are manifestly closed. The NS-NS flux through $S^3$ is of no importance here as we are only interested in the string solution that's propagating in $AdS_3$ and the string's dynamics in $S^3$ never appears in our equation of motion.   In the presence of a boundary one can incorporate a boundary  
term to parameterize
the particular ambiguity term as discussed before. For details of this ambiguity one can look at \cite{Hoare:2013lja}, where the 2-form field along the $S^3$  is written as 
$-\frac{b}{2}(\cos 2\theta + \zeta)$, with $\zeta$ is a constant that gives rises to an ambiguity term. This constant, $\zeta$ can be chosen via particular physical
requirements on the string solution itself. In our case for a string in $AdS_3$ the extra constant $\xi$ will only generate a constant shift in the desired  dispersion relation which we will see in later sections. 

To describe the motion of a rotating spiky string in this background we would follow the discussion in \cite{Kruczenski:2004wg}. The Nambu goto action in for such a string in $AdS_3\times S^3$ background with mixed flux is
  \be\label{sng}
S_{NG}=\frac{\sqrt{\l}}{2\pi}\int d\t d\s \left[\sqrt{-(\dot{X}^2X'^2-(\dot{X}.X')^2)}-\frac{\epsilon^{ab}}{2}B_{mn} \partial_a X^m \partial_b X^n\right] ,
 \ee
where $\lambda$ is the t'Hooft coupling and the antisymmetric tensor is defined accordingly $\epsilon^{01}=1$. We use the following rigidly rotating closed string ansatz to parameterise the solution,
\be
\rho=\rho(\s),\quad t=\tau ,\quad \phi=\w\tau+\s.
\ee
The equations of motion with respect to $t$ and $\phi$ can be seen to be the same, having the following form
\be\label{eom1}
\frac{\cosh^2\rho \sinh^2 \rho}{\sqrt{A}}-b\sinh^2 \rho= C
\ee
where $C$ is a integration constant and
\be\label{A}
A=\rho'^2(\cosh^2\rho -\w^2 \sinh^2 \rho)+\cosh^2\rho \sinh^2 \rho
\ee
%Note that, had we taken $B_{t\phi}=b (\sinh^2\rho +\xi)$, the above equation of motion would have read
%\be\label{eom1a}
%\frac{\cosh^2\rho \sinh^2 \rho}{\sqrt{A}}-b(\sinh^2 \rho+\xi)= C
%\ee
%The constant term $b\xi$ can be absorbed in $C$, as we would see that $C$ itself is dependent on $b$ and thus the choice of $\xi$ will not affect our calculations. Without any loss of generality, we can simply put $\xi  = 0$
% here.

Now equation (\ref{eom1}) can be used to obtain the following expression for $\rho'$ (i.e. $\sigma$ derivative of $\rho$) as 
\begin{equation}
 \rho\prime = \frac{1}{4} \frac{\sinh 2\rho}{C+ b\sinh^2\rho}\frac{\sqrt{\sinh^2 2\rho - 4(C+ b\sinh^2\rho)^2}}
 {\sqrt{\cosh^2\rho - \omega^2 \sinh^2\rho} }. \label{eom2}
\end{equation}
One can explicitly check that the above expression actually satisfies the $\rho$ equation of motion. Looking at the expression we can see that the maximum value of the string profile (the position of spikes) still goes up to $\rho_1 = \coth^{-1} \omega$ but the minimum value (position of valleys) have changed due to the inclusion of the B field. In fact, there are now two roots ($\rho_{0}^{\pm}$) corresponding to the position of valleys in the string profile. As we will see only one of them will guarantee that the angle between the cusps and the valleys remain fixed at $\frac{2\pi}{2n}$ to keep the $n$-spike string closed. 
\subsection{Spiky string profiles}

Now we make the following substitution of variables in (\ref{eom2}) \be
\cosh 2\rho = u,~~~~\cosh 2\rho_1 = u_1,~~~~\cosh 2\rho_0^{\pm} = u_0^{\pm}.
\ee
%Here the $ u_0^{\pm}$ are the roots of the equation 
%\be
%u^2(1-b^2)+u(4Cb+2b^2)-(1+b^2+4C^2+4Cb) = 0.
%\ee
Using the above mentioned change of variables, we can write the expression for $\sigma$ as the following integral,
\begin{equation}\label{sigma1}
\int d \sigma = \int\frac{b\sqrt{\omega^2 -1}}{\sqrt{2}\sqrt{1-b^2}}\frac{u+\frac{2C}{b}-1}{u^2-1}
 \frac{\sqrt{u_1- u}~~du}{\sqrt{u^2(1-b^2)+u(4Cb+2b^2)-(1+b^2+4C^2+4Cb)}}
\end{equation}
The term $u^2(1-b^2)+u(4Cb+2b^2)-(1+b^2+4C^2+4Cb)$ in the above equation is quadratic in $u$ and hence can be decomposed in the form $(u-u_0^{+})(u-u_0^{-})$, where  $ u_0^{\pm}$ are the roots.
%\begin{equation}
% d\sigma = \frac{b\sqrt{\omega^2 -1}}{\sqrt{2}\sqrt{1-b^2}}\frac{u+\frac{2C}{b}-1}{u^2-1}
% \frac{\sqrt{u_1- u}}{\sqrt{(u-u_0^{+})(u-u_0^{-})}}
%\end{equation}

Here the full algebraic form of the roots can be written as,  
\begin{equation}
 u_1 = \frac{\omega^2+1}{\omega^2-1} \label{long}
\end{equation}
and corresponding to the valleys we have,
\begin{equation}
 u_0^{\pm} = \frac{b^2-2Cb\pm \sqrt{1+4C^2-4Cb}}{-1+b^2}.
\end{equation}
Here if one looks closely, only the root $u_0^{-}$ will lead to the original $u_0$
for the $b=0$ case as explained in \cite{Kruczenski:2004wg}. So this will be our root of interest, i.e. where the minimum positions of the string occurs. This can also be justified by the form of the total integral,
\begin{equation}
 \sigma=  -\frac{b\sqrt{\omega^2 -1}}{\sqrt{2}\sqrt{1-b^2}\sqrt{u_1-u_0^{+}}}
 \left[2\mathbf{F}(\alpha,\beta)-\frac{2C}{b}\mathbf{\Pi}(\gamma_{-},\alpha, \beta)+
 (\frac{2C}{b}-2)\mathbf{\Pi}(\gamma_{+},\alpha, \beta)\right], \label{profile}
\end{equation}
Where $\mathbf{F}$ and $\mathbf{\Pi}$ are the standard incomplete elliptic integrals of second and third kind. Here we also have the arguments as
\begin{equation}
 \sin\alpha = \sqrt{\frac{u_1-u}{u_1-u_0^{-}}}~~~~~\beta = \frac{u_1-u_0^{-}}{u_1-u_0^{+}}~~~~~
 \gamma_{\pm} = \frac{u_1-u_0^{-}}{u_1\pm 1}.
\end{equation}
 The angle difference between the spikes and $(\rho=\rho_1)$ the valleys $(\rho=\rho_0)$ is $\Delta \phi = \frac{2\pi}{2n}$ can be obtained from (\ref{profile}) as follows
\begin{equation}\label{dphif}
 \Delta\phi=  -\frac{b\sqrt{\omega^2 -1}}{\sqrt{2}\sqrt{1-b^2}\sqrt{u_1-u_0^{+}}}
 \left[2\mathbf{F}(\frac{\pi}{2},\beta)-\frac{2C}{b}\mathbf{\Pi}(\gamma_{-},\frac{\pi}{2}, \beta)+
 (\frac{2C}{b}-2)\mathbf{\Pi}(\gamma_{+},\frac{\pi}{2}, \beta)\right].
\end{equation}
\begin{figure}[t]
        \centering

                \includegraphics[width=1.02\linewidth]{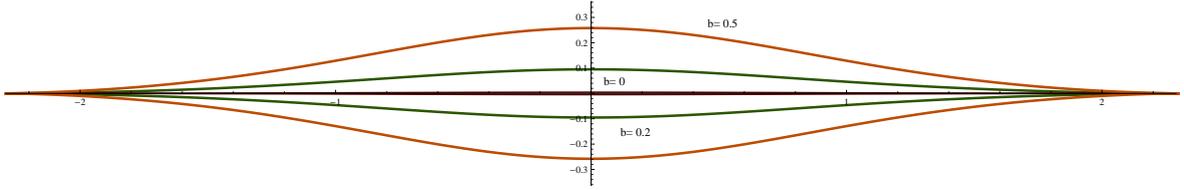}

        \caption{Closed spiky string profiles with $n=2$ for various values of the flux parameter $b$. The $b= 0$ case corresponds to that of the folded string, which is a straight line through the center. Here we have taken $\omega= 1.02$.  Notice the `fattening' of the string as we keep increasing $b$. The string profiles having non-zero $b$ does not actually pass through the center anymore.}\label{fig:c}
\end{figure}        

\begin{figure}[t]
        \centering

                \includegraphics[width=0.85\linewidth]{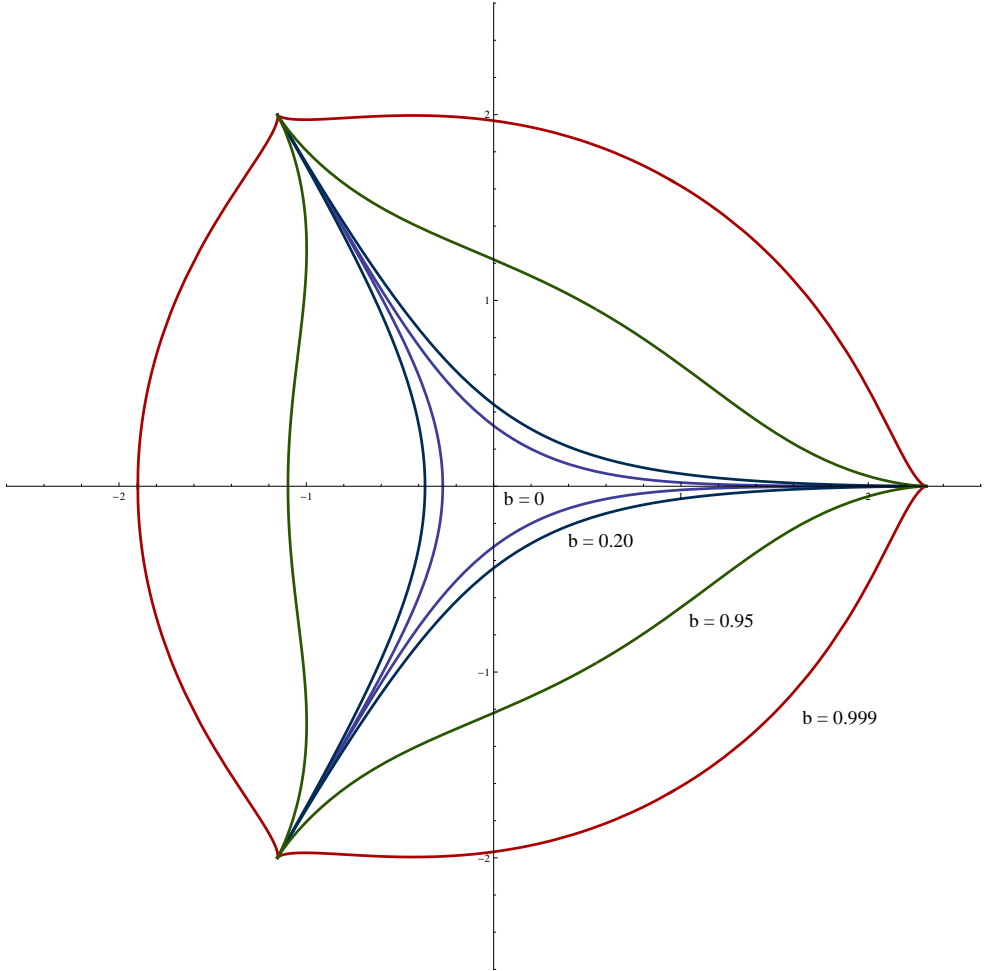}

        \caption{Closed spiky string profiles with $n=3$ for various values of the flux parameter $b$. We have taken $\omega= 1.02$ and fixed the values of $C$ accordingly. Notice the `fattening' of the string with increasing $b$.}\label{fig:a}
\end{figure}

\begin{figure}[t]
        \centering

                \includegraphics[width=0.85\linewidth]{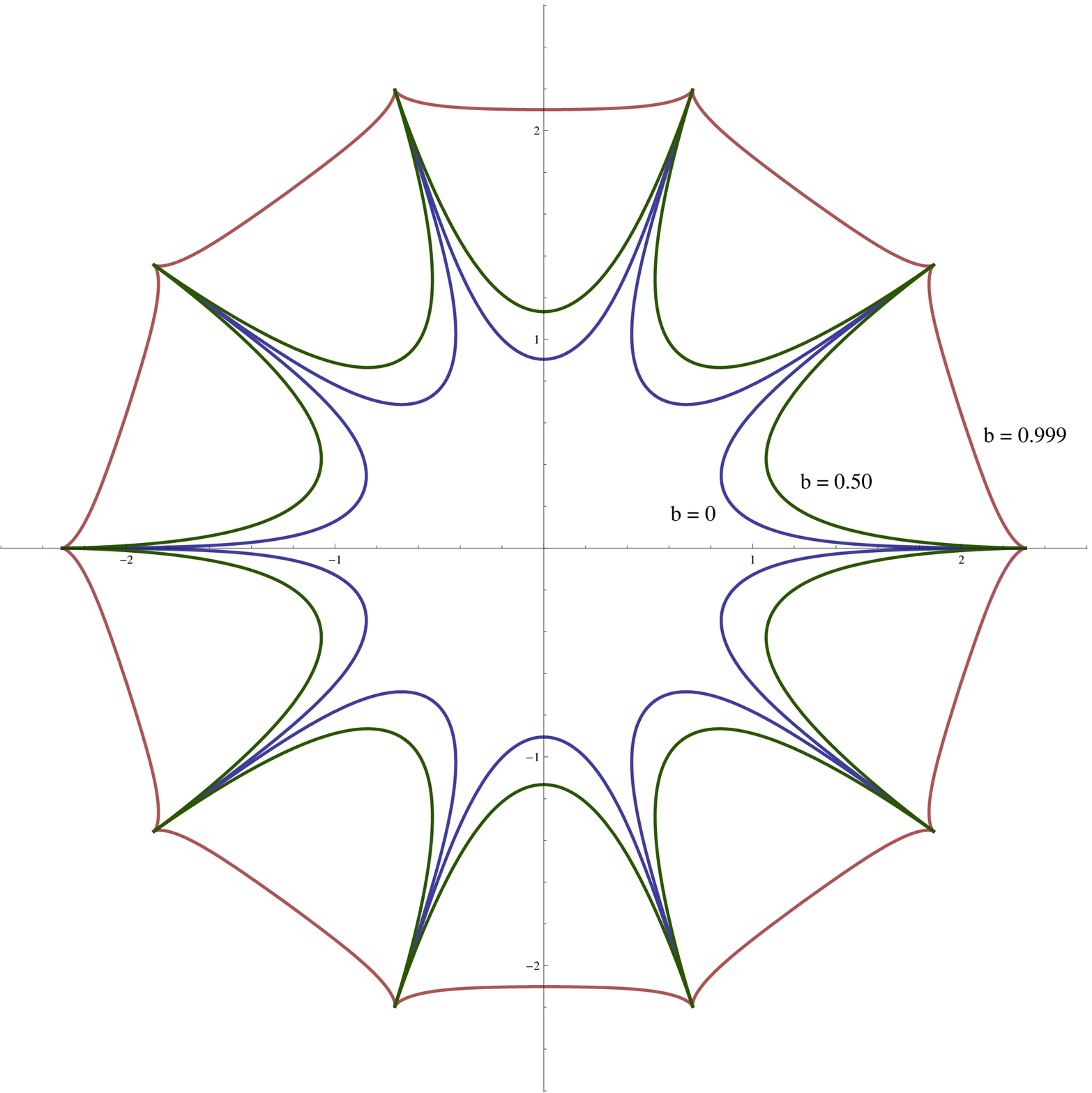}

        \caption{Closed spiky string profiles with $n=10$ for various values of the flux parameter $b$. We have taken $\omega= 1.02$ and fixed the values of $C$ accordingly. Notice the `fattening' of the string with increasing NS-NS flux.}\label{fig:b}
\end{figure}

This will determine the total spiky string profile for our case. It is worth noting that by fixing $\omega$ we can fix the spike height $\rho_1$, and putting in different values of $b$ we can recover the corresponding values of $C$ by implementing the closedness condition $\Delta\phi= \frac{2\pi}{2n}$,  $n\in \mathbb{Z}$. By this procedure, we fix all the parameters in the calculation. In figures \ref{fig:c}, \ref{fig:a} and \ref{fig:b} we present the polar plot of the string profiles for $n=2$, $n=3$ and $n=10$ from equation (\ref{profile}) and the closedness condition on the string for different values of $b$. We should remember at this point that $n=2$ and $b=0$ corresponds to the GKP folded string in this case. Now it is clear from (\ref{long}) that the `long' string limit ($\rho_1\rightarrow \infty$) on the solution can be taken when $\omega \rightarrow 1$. We take the values of $\omega$ accordingly. It is worth noting from the figures that as we increase the value of $b$ the string becomes fatter, and in the limit $b\rightarrow 1$ it almost approaches the shape of a circular string, although the cusps remain intact. $b=1$ is a special limit in which RR flux vanishes and the string profile cannot be continuously obtained from (\ref{profile}).   This case is discussed in a bit more details in section \ref{pure}.

\subsection{Conserved charges}
Here, we can write the (\ref{eom2}) in the following suggestive form
 \be 
\rho'^2=\frac{\cosh^2\rho \sinh^2 \rho}{\cosh^2 \rho-\w^2 \sinh^2 \rho}\left[\frac{\cosh^2\rho \sinh^2 \rho -(C+b \sinh^2\rho)^2}{(C+b \sinh^2\rho)^2}\right]. \label{eom4}
\ee
%where $B=C+b \sinh^2\rho$.
 To make the concept of cusps and valleys clearer, we demand that $\rho '$ vanishes for a particular value $ \rho=\rho_0$ as was done in \cite{Kruczenski:2004wg}. Thus, $C$ can be written in terms of $\rho_0$ by choosing the appropriate root as in the section before,
\be
C=\sinh\rho_0 \cosh \rho_0-b\sinh^2\rho_0. \label{valley}
\ee
This makes the expressions more tractable 
 and the integration constant reduces to the correct one for the pure RR case. Instead of $\rho_0^-$ or equivalently $u_0^-$, we would use $\rho_0$ itself as the position of the valleys hereafter in this paper. The equation of motion for $\rho$ can be inverted to write the closedness condition on the $n$-spike string,
\be \label{sigma}
\int \frac{d\rho}{\rho'}=\int d\s
\ee
where $\rho'$ is as defined in (\ref{sigma1}).
%After changing the variable $\cosh2\rho =u$, $\cosh2\rho_1 =u_1$ and $\cosh2\rho_0 =u_0$, the integral can be written as 
%
%\begin{eqnarray}
%\frac{\left(2c-b\right)}{\sqrt{2}\sinh \rho_1}\int \frac{\sqrt{u_1-u}\,du}{(u^2-1)\sqrt{X}}+\frac{b}{\sqrt{2}\sinh \rho_1}\int \frac{\sqrt{u_1-u}\,u\,du}{(u^2-1)\sqrt{X}}=d\s
%\end{eqnarray}
%
%where 
%
%\begin{eqnarray}\label{ropol}
%X&=&u^2-1-\left[\sqrt{u_0^2-1}+b(u-u_0)\right]^2\\ \nn
%&=&(1-b^2)\left(u^2-2ub\frac{(\sqrt{u_0^2-1}-bu)}{1-b^2}-\frac{(\sqrt{u_0^2-1}-bu_0)^2+1}{1-b^2}\right)\\ \nn
%&=&(1-b^2)\left[(u-u_+)(u-u_-)\right]
%\end{eqnarray}
%where $u_+$ and $u_-$ are the roots of the polynomial in the second line of \ref{ropol}.
%\be\label{roots}
%u_+=u_0,\quad u_-=\frac{u_0+b^2 u_0-2 b \sqrt{u_0^2-1}}{b^2-1}
%\ee
Using the isometries of the background, we can construct the Noether charges. The expression for the spin of the string, $S$ is as follows\be\label{spin0}
S=\frac{2n\sqrt{\l}}{2\pi}\frac{\w}{2}\int\tanh \rho \sqrt{\frac{\sinh ^2\rho \cosh ^2\rho-(C+b \sinh^2\rho)^2}{\cosh ^2\rho -\omega ^2 \sinh ^2\rho }}d\rho =\frac{n\sqrt{\l}\w}{\pi}{\cal S},
\ee
where $\cal{S}$ is the rescaled spin defined for simplifying the expressions.
To find the desired dispersion relation we write the following expression for $E-\w S$
\be
E-\w S=\frac{2n\sqrt{\l}}{2\pi}\int \left[\frac{\cosh^2\rho \rho'^2}{\sqrt{A}}+\frac{\cosh^2\rho\sinh^2\rho}{\sqrt{A}}-b\sinh^2\rho-\frac{\w^2\sinh^2\rho \rho'^2}{\sqrt{A}}\right]d\s,
\ee
where $A$ is given by (\ref{A}).
Implementing the eom (\ref{eom1}) and simplifying a bit, we obtain
\begin{eqnarray}\label{disps}
E-\w S=\frac{2n\sqrt{\l}}{2\pi}\int \Bigg[\sinh \rho  \cosh \rho  \sqrt{\frac{\cosh ^2\rho -\omega ^2 \sinh ^2\rho }{\sinh ^2\rho  \cosh ^2\rho -(C+b \sinh^2\rho)^2}}\\ \nn
-b( C+b \sinh^2\rho) \tanh \rho  \sqrt{\frac{\cosh ^2\rho -\omega ^2 \sinh ^2\rho }{\sinh ^2\rho  \cosh ^2\rho -(C+b \sinh^2\rho)^2}}\Bigg]d\rho.
\end{eqnarray}

%Another way to express it in terms of $u$ is
%\be
%E-\w S=\frac{2n\sqrt{\l}}{2\pi}  \left(c\int d\s+\frac{\sqrt{1-b^2}}{2 \sqrt{2}\sinh \rho_1}\int \frac{\sqrt{(u-u_+)(u-u_-)(u_1-u)}}{u^2-1}du\right)
%\ee
%where $u_+$ and $u_-$ are the roots in (\ref{roots})
%Note that this way of writing the equation is different from Kruczensky.
%The expression of the angular momentum can similarly be written as,

\subsection{Perturbation theory in $b$ and dispersion relations}

Since the integrals corresponding to the charges mentioned in the last subsection are quite complicated to evaluate, to find the dispersion relation we expand the integral expressions around small $b$ up to the order of $b^2$ and substituting $\w=1$. This refers to the case when amount of NS-NS flux is small and the spikes are `long' enough to approach the boundary of $AdS$. The final dispersion relation will also be evaluated perturbatively in orders of small $b$ upto ${\cal O}(b^2)$. 
The expression for the angle difference $\Delta \phi$ can be written as,
\begin{eqnarray}\label{dphipert}
\Delta \phi &=& \int_{u_0}^{u_1}\left[\frac{\sqrt{u_0^2-1}}{(u^2-1)\sqrt{u^2-u_0^2}}+\frac{b }{(u+u_0)\sqrt{u^2-u_0^2}}+\frac{3b^2\sqrt{u_0^2-1}}{2(u+u_0)^2\sqrt{u^2-u_0^2}}\right]du+{\cal O}(b^3)~~~~ ~~~~\\ \nonumber
&=&\frac{\pi}{2}-\sin^{-1}\frac{u_1 \sqrt{u_0^2-1}}{u_0\sqrt{u_1^2-1}}+\frac{b \sqrt{u_1-u_0}}{u_0 \sqrt{u_1+u_0}}+\frac{b^2 \sqrt{u_0^2-1} \sqrt{u_1-u_0} (u_1+2 u_0)}{2 u_0^2 (u_1+u_0)^{3/2}} +{\cal O}(b^3)
\end{eqnarray}
The above equation can also be obtained by performing perturbation of (\ref{dphif}) in orders of $b$. We will drop ${\cal O}(b^3)$ from our subsequent equations of this section but it will be implied that these are written upto ${\cal O}(b^2)$.
 In the limit $u_1>>u_0$ (i.e. the long string limit) it can be seen from the second line of (\ref{dphipert}) that $u_1$ dependence is canceled off in the expression for $\Delta \phi$. Thus after inverting the equation $u_0$ will only be a function of $\Delta \phi$ in this limit. The expression for $u_0$ in the small $b$ regime will be given by,
\be\label{rho0f}
u_0=\frac{1}{\sin (\Delta \phi )}+b \cot (\Delta \phi )+\frac{b^2}{2 \sin (\Delta \phi )} 
\ee
%The expression for the spin $S$ can be written in a rescaled version to define a new quantity,
%\be
% J=\frac{\pi S}{\sqrt{\l}n\w}
%\ee
%where we can write the integral form of $J$ upto $\mathcal{O}(b^2)$ as follows
Similarly the perturbative expression for spin ${\cal S}$ (\ref{spin0}) can be written by expanding it upto $\mathcal{O}(b^2)$ as follows
\begin{eqnarray}
{\cal S}= \int_{\rho_0}^{\rho_1}&& \Bigg[ \frac{1}{2} \tanh \rho  \sqrt{\frac{\sinh^2 2 \rho -\sinh^2 2 \rho_0}{ \cosh ^2\rho - \omega ^2 \sinh ^2\rho }}\\ \nn
&&-\frac{b \tanh \rho  \sinh 2 \rho_0 \sqrt{\frac{2\sinh^2 2 \rho -2\sinh^2 2 \rho_0}{-\left(\omega ^2-1\right) \cosh 2 \rho +\omega ^2+1}}}{2 (\cosh 2 \rho +\cosh 2 \rho_0)}-\frac{b^2 \sinh ^3\rho  \cosh \rho  \sqrt{\frac{2\sinh^2 2 \rho -2\sinh^2 2 \rho_0}{-\left(\omega ^2-1\right) \cosh 2 \rho +\omega ^2+1}}}{(\cosh 2 \rho +\cosh 2 \rho_0)^2}\Bigg]d\rho. \nn
\end{eqnarray}

 After evaluating this integral with $\w=1$ and using the limit $\rho_1 >> \rho_0$, the resulting equation can be inverted in the orders of $b$ to obtain $\rho_1$ in terms of $\cal{S}$ and $\rho_0$ or $u_0$.
\be\label{rho1f}
\rho_1=\frac{1}{2} \log (8 \,{\cal S}) +\frac{b \sqrt{u_0^2-1}}{8  \, {\cal S}} \log (8 \, {\cal S})+\frac{b^2 \left[16\, {\cal S}^2-\left(u_0^2-1\right) (\log (8\, {\cal S})-2) \log (8\, {\cal S})\right]}{64\, {\cal S}^2}
\ee

The expression for the dispersion relation can be written in terms of $u$ and it is as follows,
\begin{eqnarray}
E-  S=\frac{2n\sqrt{\l}}{\pi} \int_{u_0}^{u_1} \Bigg[ \frac{1}{2\sqrt{u^2-u_0^2}}-\frac{b (u_0-1)^{3/2} \sqrt{u_0+1} }{2 (u+1) (u+u_0)\sqrt{u^2-u_0^2}}\\ \nn
-\frac{b^2  \left(u^2+2 u (u_0-1)-2 u_0 (u_0+1)+3\right)}{4 (u+u_0)^2\sqrt{u^2-u_0^2}}\Bigg] du
\end{eqnarray}

After performing the above integral $\rho_1$ and $\rho_0$ can be substituted from (\ref{rho0f}) and (\ref{rho1f}) respectively and the final expression will be in terms of $E$, $S$ and $\Delta \phi=\frac{2\pi}{2n}$. After some algebra, the desired dispersion relation in the limit $S>>\sqrt{\l}$ can be written as,
\begin{eqnarray}\label{finaldisp}
E-S=\frac{\sqrt{\l}n}{\pi}\Bigg[\frac{1}{2}\log S+{\cal O} \left(\frac{1}{S}\right) &+&b\left(\cos \frac{\pi}{n}-\frac{1}{2}\cos^{-1}\sin \frac{\pi}{n}+{\cal O} \left(\frac{1}{S}\right)\right)\\ \nn
&&-\frac{b^2}{4}\left(\log S+{\cal O} \left(\frac{1}{S}\right)\right)\Bigg]+{\cal O}(b^3). 
\end{eqnarray}
We note that there is no $(\log S)^2$ term present in the leading order as was for the folded string solution. Note that the leading term in the ${\cal O}( b)$ term is a constant.  For $n=2$, $\Delta \phi=\frac{\pi}{2}$, the ${\cal O}(b)$ terms in the expression for $\rho_0$, $\rho_1$ and $E-S$ cancel out and we get the following dispersion relation
\be 
E-S=\frac{\sqrt{\l}}{\pi}\left[ \log S-\frac{b^2}{2}\log S\right]+{\cal O}(b^3). \label{disp}
\ee
As usual, putting $b=0$ in the above expression gives us back the simple dispersion relation of the straight `long' GKP string. But without even having a relation to the $b \neq 0$ folded string, the above expression generalises the $n=2$, $b=0$ spiky string in this background. Note that while we could not find the exact $b$ dispersion relation using our method, the one presented here looks very natural. It can also be commented that in general, the relation (\ref{finaldisp}) looks to be a small $b$ expansion of the general relation $E - S \sim \sqrt{1-b^2}\log S + b. \text{Const}$. Although we can't but speculate about this since we don't have enough information to make the claim.

Another interesting thing that can be mentioned here is the ${\cal O}(b)$  term present in (\ref{finaldisp})  has an exact form that cancels out in the limit $n=2$. For example, had we chosen the NS-NS flux including the ambiguity term in the form $b (\sinh^2\rho +\xi)$, then the small $b$ expansion of the $E-S$ integral would actually include another constant term proportional to $b\xi$ as we had speculated earlier. Therefore, the choice of $\xi = 0$ explicitly gives rise to a constant term that vanishes when $n=2$.

\subsection{Notes on the infinite spin limit}
In the infinite spin limit the spikes of the string actually touch the $AdS$ boundary. This particular limit, as we have shown, corresponds to $\omega \to  1$ without depending upon whether the NS-NS flux is included or not. In this limit the shape of the string simplifies and for the $b=0$ case it has been discussed in details at various places including in \cite{Kruczenski:2008bs} . We can briefly mention the results here and try to compare them with expressions in our case.

We consider the $\rho$ equation of motion for the $b=0$ case as outlined in \cite{Kruczenski:2004wg} ,
\be
\rho\prime = \frac{1}{2} \frac{\sinh 2\rho}{\sinh 2\rho_0^{(0)}}\frac{\sqrt{\sinh^2 2\rho - \sinh^2 2\rho_0^{(0)}}}
 {\sqrt{\cosh^2\rho - \omega^2 \sinh^2\rho} }. \label{eom3} 
\ee
Here the $\rho_0^{(0)}$ indicates the position for `valleys' in the string without turning on any NS-NS flux. In the large spin limit, we can put $\omega = 1$ in the above and integrate it to get the string profile as,
\be\label{n1}
\tan(\sigma +c) = \frac{\sinh 2\rho_0^{(0)} \cosh 2\rho}{\sqrt{\cosh^2 2\rho -\cosh^2 2\rho_0^{(0)} }}.
\ee
Choosing $c = \frac{\pi}{2}$, we ensure that $\rho = \rho_0^{(0)}$ at $\sigma = 0$ is the position where the `valleys' lie. At the tip of the spikes in this case we have $\rho \to \infty$, which corresponds to a value $\pm\sigma_c$. It is then straightforward to understand that the angular separation between two spikes can be given by,
\be
2\Delta \phi = 2\sigma_c= 2\frac{\pi}{n}.
\ee
It can now be shown using the above conditions that with $\rho \to \infty$, we can write
\be
\cot \sigma_c = \sinh 2\rho_0^{(0)} \label{regularity}.
\ee
It can be clearly seen here that at $n=2$, $\sigma_c = \frac{\pi}{2}$, i.e. at the limit where the spiky string should become a straight folded string hinged at $\rho_0^{(0)} = 0$. This of course corresponds to the string passing through the center of $AdS$ with ends reaching upto the boundary $(\rho=\infty)$ as we would expect.  The string profile can be written in the form,
\be\label{n2}
\coth 2\rho = \frac{\cos \sigma}{\cos \sigma_c}.
\ee
 
 But this $n=2$ limit can't be taken very naively as the spiky string and the folded string are really very different type of strings as can be emphasised by the ansatz used to parameterize them. While for the folded string \cite{Gubser:2002tv}, the angular direction $\phi = \omega \tau$, the spiky string \cite{Kruczenski:2004wg} $\phi$ is linear in $\sigma$. But one must remember this problem goes away as the valleys of the spiky string ($\sigma = 0$) actually becomes the center of the folded string in the $n=2$ limit. One other way to visualise the limit, which has been extensively discussed in \cite{Kruczenski:2008bs}, is to perform a conformal transformation to one arc of the spiky string to relate it to the straight folded string. In the large spin limit the arc between two spike tips correspond to a wilson loop surface ending on two parallel lightlike lines on the boundary. In this limit, using the $SO(2,2)$ symmetry of the $AdS_3$ we can transform a generic arc of the spiky string to the folded string which in turn corresponds to $\sigma_c = \frac{\pi}{2}$ or $\rho_0^{(0)}= 0$.
 
 For our case, the inclusion of the NS-NS flux makes the problem quite complex indeed. Let us start from (\ref{eom4}) and take the infinite spin limit of the spiky string by putting in $\omega = 1$. We integrate the equation to find,
 \be\label{n3}
\tan( \sigma + \tilde{c}) =\frac{2\sqrt{1-b^2}}{\sinh 2\rho_0}\frac{\frac{\sqrt{\cosh 2\rho - \cosh 2\rho_0}}{\sqrt{\cosh 2\rho - \cosh 2\tilde{\rho_0}}}}{(1-b^2)-(\coth\rho_0-b)(b-\tanh\rho_0)\frac{(\cosh 2\rho - \cosh 2\rho_0)}{(\cosh 2\rho - \cosh 2\tilde{\rho_0})}}.
 \ee
 In the above equation, we have  used the definition of $\tilde{\rho_0}$
 \be 
 \cosh2\tilde{\rho_0} = \frac{2b}{1-b^2}\sinh 2 \rho_0 -\frac{1+b^2}{1-b^2}\cosh 2 \rho_0.
 \ee

 And $\rho_0$ is the position of the `valley' as discussed before in (\ref{valley}). $\rho_0$ isn't to be confused with the $\rho_0^{(0)}$ in the $b=0$ case as it is now dependent on $b$ itself. Although the positions of the `valleys' reduce to the the earlier values as we put $b=0$.  We can choose $\tilde{c}= 0$ which implements the condition that at $\rho = \rho_0$, $\sigma =0$ as in the case of no flux turned on. With this choice of $\tilde{c}$ (\ref{n3}) reduces to (\ref{n1}) for $b=0$, as expected. Similarly as the spike  tips are now touching the boundary, we want that at $\rho \to \infty$, $\sigma \to \pm \sigma_c = \pm\frac{\pi}{n}$. Using this condition, after a little algebra, we obtain the result as
 \be
 \cot \sigma_c = \frac{1}{\sqrt{1-b^2}}(\sinh 2 \rho_0 - b\cosh 2\rho_0). \label{regularity2}
 \ee
 This is an \textit{exact} relation in $b$, valid for the large spin limit. Comparing the above with (\ref{regularity}) we can see the stark difference due to the inclusion of the NS-NS flux. In this case, at $n=2$ or $\sigma_c=\frac{\pi}{2}$, $\rho_0  = \frac{1}{2} \tanh^{-1} b$, i.e. the analogue of $n=2$ spiky string in this case is not actually like a straight folded string and  does no pass through the center of $AdS$. This exact fattening of the string was visualised in Figure (\ref{fig:c}) as it can be seen that the string profiles stop passing through $\rho = 0$ as we increase $b$. We can explicitly calculate to see that the  integration constant $C=\frac{b}{2}$ for this value. As expected we can get (\ref{regularity}) back by simply putting $b=0$ into (\ref{regularity2}). 
 
The equation (\ref{n3}) looks quite involved as opposed to the $b=0$ case and is much difficult to handle. We refrain from going into further quantitative details on the shape of the string here.

 \section{The case of pure NS-NS flux}\label{pure}

The special case of $b=1$ corresponds to the scenario where the $AdS_3 \times S^3$ is supported purely by NS-NS flux and there is no R-R flux. In this limit, as we have mentioned before, the theory can be described by a $SL(2,\mathbb{R})$ WZW model \cite{Maldacena:2000hw}. Strings with large spin $S$ in this background were also studied in \cite{Loewy:2002gf}.

The equation (\ref{sigma}) simplifies considerably in the limit $b=1$ and $\w=1$. We obtain exact trigonometric relations instead of elliptic integrals. We have set $\w$  to $1$ to obtain the spikes at infinity.

\be\label{b1s}
\s=\tan^{-1} \frac{\sqrt{2} \left(\coth \rho_0^{(1)}+1\right) \text{csch}\rho_0\, \sqrt{\cosh 2 \rho -\cosh 2 \rho_0^{(1)}}}{(\coth \rho_0^{(1)} +1)^2-\cosh 2 \rho \,\, \text{csch}^2\rho_0^{(1)}},
\ee
%\be
%\rho=\cosh ^{-1}\left(\frac{e^{2 \text{$\rho $0}} \left(-\sqrt{\left(1-e^{-4 \text{$\rho $0}}\right) \text{T$\sigma $}^2+1}+\text{T$\sigma $}^2+1\right)}{\text{T$\sigma $}^2}\right)
%\ee
where at $\s=0 ,\,  \rho=\rho_0^{(1)}$ boundary condition has been imposed to evaluate the integral.

 The equation for the angular difference $\Delta \phi$ from valley $(\rho=\rho_0^{(1)})$ to peak$(\rho=\rho_1)$ in the limit $\rho_1>>\rho_0>>1$ can be obtained very easily from (\ref{b1s})
\be\label{dp1}
\tan \Delta \phi=  \tan \frac{\pi}{n} \simeq 2e^{-(\rho_1-\rho_0^{(1)})}.
\ee
Since for a long string $\rho_1\rightarrow \infty$, from the above equation we can conclude the number of spikes($n$) for long spikes has to be large (${\cal O}(e^{\rho_1})$). The expression for spin(\ref{spin0}) in this limit can be evaluated as follows,
\be\label{spin1}
S\simeq \frac{n \sqrt{\l}}{2\pi}e^{(\rho_1-\rho_0^{(1)})}.
\ee
%
%\be
%\frac{\tan \Delta \phi}{2} \simeq e^{-\rho_1}\simeq \frac{n \sqrt{\l}}{2\pi S} \quad  \Rightarrow \quad S\simeq \frac{n^2\sqrt{\l}}{\pi^2}
%\ee
% 
Note that in this limit, the spin, $S$ and the number of spikes, $n$ are not independent of each other and can be related using the equations (\ref{dp1}) and (\ref{spin1}
\be
S=\frac{n\sqrt{\l}}{2\pi}\frac{2}{\tan \frac{\pi}{n}}\simeq \frac{n^2 \sqrt{\l}}{\pi^2},
\ee
where in the last line we have used $\tan \frac{\pi}{n}\simeq \frac{\pi}{n}$ for large $n$.

The dispersion relation, (\ref{disps}) in the above mentioned limit takes the following unique form
\be
E-S=\frac{n\sqrt{\l}}{\pi} \tan ^{-1}e^{(\rho_1-\rho_0^{(1)})}=\frac{n\sqrt{\l}}{\pi} \tan ^{-1}\frac{2\pi S}{n \sqrt{\l}}. \label{disp2}
\ee
The above dispersion relation looks completely different from the one for the usual spiky strings. One has to understand this relation in a better way.

%In the limit $\w=1$, these integrals can be solved to obtain
%
%\be
%\Delta \phi = \frac{\pi}{2}-\sin^{-1}\left( \frac{u_1\sqrt{u_0^2-1}}{u_0 \sqrt{u_1^2-1}}\right)+\frac{b \sqrt{u_1-u_0}}{u_0 \sqrt{u+u_0}}+\frac{b^2 \sqrt{u_0^2-1} \sqrt{u_1-u_0} (u_1+2 u_0)}{2 u_0^2 (u_1+u_0)^{3/2}}
%\ee
%
%To perform the $S$ integral we need to get rid of the $\rho_0$ terms in the integral itself following the arguments in Kruczensky's paper.
%
%\begin{eqnarray}
%S&=&\frac{2n\sqrt{\l}}{2\pi}\Big(\frac{ \sinh (2 \rho ) \tanh (\rho )}{2 }-\frac{1}{2} b \tanh (\rho ) \tanh (2 \rho ) \sinh (2 \rho_0)\\ \nn
%&&-b^2 \sinh ^3(\rho ) \cosh (\rho ) \tanh (2 \rho ) \text{sech}(2 \rho )\Big) \\ \nn
%&=&\frac{2n\sqrt{\l}}{2\pi}\Big(\frac{1}{4} \sinh (2 \rho )-\frac{\rho }{2}-b \sinh (2 \rho_0) \left(\frac{\rho }{2}-\frac{1}{2} \tan ^{-1}(\tanh (\rho ))\right) \\ \nn
%&&-\frac{1}{8} b^2 \left(\sinh (2 \rho )+\tanh (2 \rho )-2 \left(\rho +\tan ^{-1}(\tanh (\rho ))\right)\right)\Big) \\ \nn
%&=& \frac{2n\sqrt{\l}}{2\pi}\Big(\frac{e^{2\rho_1}}{8}-b \frac{\sinh (2\rho_0)\rho_1}{2}-b^2\frac{e^{2\rho_1}}{16}\Big)
%\end{eqnarray}
%
%\begin{eqnarray}
%E-S=\frac{2n\sqrt{\l}}{2\pi} \Bigg(\frac{1}{2}\cosh^{-1}\frac{u_1}{u_0}+O(b)+O(b^2)\Bigg)
%\end{eqnarray}
%The $O(b)$ terms seems to be imaginary! This can have some non trivial physical implications. Like it can be interpreted as a trivial phase if the dispersion relation is a scattering amplitude of some sort! 
%
\section{Conclusions and discussions} \label{conclusions}
In this paper we have studied the spiky spinning string in the $AdS_3\times S^3$ background supported by mixed Ramond Ramond and Neveu Schwarz fluxes. Our main result can be summarized by eq. (\ref{finaldisp}). The case of $n=2$ spikes is interesting since for a pure RR background, the $n=2$ spiky string resembles the long folded string \cite{Kruczenski:2004wg}. From our analysis of linear perturbation in NS-NS flux coefficient we have found that the dispersion relation of the $n=2$ spiky string gets corrected by a factor of $-\frac{b^2}{2\pi}\log S$ in addition to the usual $\frac{1}{\pi}\log S$. However, this result is different from that of the long folded string of \cite{David:2014qta} where the leading correction was found out to be $-\frac{b^2}{2\pi}(\log S)^2$.  
%Contradicting known literature, in this case we find that the dispersion relation for spiky strings does not reduce to that of folded string solution in the limit of $n=2$.
 
 The physical explanation for this fact is not entirely clear to us. It might be the case that under presence of NS fluxes, folded strings and 2-spike strings might actually correspond to completely different objects in the dual CFT. We haven't been able to reproduce the string solution of \cite{David:2014qta} from our spiky string through any series of SO(2,2) reparametrization of the string solution in $AdS_3$. The other important thing to note here is that our construction of the spiky string closes exactly  in contrast to the solution in \cite{David:2014qta} where a large number of windings were necessary to close the string. Also, in \cite{Banerjee:2015bia}, it was noted that a large number of windings are needed for a circular string in $AdS_3$ to maintain pulsating motion. In the present case, such conditions do not come into the picture, which might indicate that the 2-spike string and the folded one are actually different object even from the string point of view. 
 
 One of the other inputs in this case comes from the idea of Wilson loops in this background. As mentioned earlier, in $AdS_5/CFT_4$ correspondence, the dispersion relation of spiky string is related to the anomalous dimension $f(\l)$ of twist two operators, for large spin, $S$ in the planar limit of $
{\cal N}=4$ theory in 4 dimension,
\be \label{anom}
\Delta-S=f(\l)\log S
\ee

Now, $f(\l)$ can also be read off from the coefficient of the logarithmic divergence of a Wilson loop which makes a cusp of angle $\gamma$ \cite{Kruczenski:2002fb}. When the cusp angle is large, the expectation value is given by
\be
W=\left( \frac{L}{\epsilon}\right) ^{f(\l)\g}
\ee
 The coefficient $f(\l)$ for the case of $AdS_3\times S^3$ background with mixed three form fluxes was calculated in \cite{David:2014qta}. To  achieve this the authors constructed minimal surfaces which end on a light like cusp in the Poincare coordinates
of $AdS_3$ with mixed three form flux. The expectation value of the Wilson line was classically found out to be(see Eq(3.13) of \cite{David:2014qta})
\begin{equation}\label{wilson}
 \langle W\rangle = \exp({-S_{NG} }) = 
 \left( \frac{L}{\epsilon}\right) ^{-\frac{\sqrt{\l}}{2\pi} \left[\frac{1}{2}\pm i \frac{b}{2}-\frac{b^2}{4}+O(b^3)\right]\g}. 
\end{equation}
Since the imaginary term just contributes to the phase, the correction to $f(\l)$ is a relative factor $-\frac{b^2}{2}$. We speculate that the anomalous dimension of cusped Wilson loop $f(\l)$ captures the coefficient of $\log S$ instead of $(\log S)^2$ as can be seen from (\ref{anom}).  This agrees with our result of dispersion relation for $n=2$ spiky strings (\ref{disp}). However, we also would like to comment that since little is known about the dual CFT, we are not sure if the equation (\ref{anom}) would hold true for twist 2 operators in this setting. It is not entirely clear here what would be the correct generalization of the GKP \cite{Gubser:2002tv} string or the string corresponding to the twist two operator in $AdS_3$ background with NS-NS flux. In our case, the $n=2$ spiky string looks somewhat better suited for this cause as upto the  $\mathcal{O}(b^2)$ we can't find a contribution that diverges faster than $\log S$ in the large spin limit. But again, we have shown that the $n=2$ strings in this case is not like the usual folded straight string as the `fatness' of the profile is being regulated by the amount of NS-NS flux turned on. More investigations from the string side are needed to shed light on this matter.

There are a number of problems that can be studied in conjunction to this. For example it would be interesting to include an angular momentum along $S^1$ into the solution and study the spiky strings as done in \cite{Ishizeki:2008tx}. It would also be great to study fluctuations around such string solutions to evaluate the effect of the NS flux. Moreover, as noted previously, looking at (\ref{disp}) one might wonder if one can find a dispersion relation for such a string exact in the flux parameter $b$, which remains valid in the total $1<b<0$ regime. We could not find such an exact relation from our construction. One way to do that might be to start with the $b = 0$ solution and construct the $b \neq 0$ solution via worldsheet transformations as done for the Giant Magnon case in \cite{Hoare:2013lja}. Of course, the implications of the string dispersion relation for pure NS flux (\ref{disp2}) remains mysterious in this case. We hope to address some of these issues in near future.

\section*{Acknowledgements}
AB would like to thank Kamal L Panigrahi for his constant support and encouragement during this work. It is a pleasure to thank Ashoke Sen, Justin R David, Anirban Basu, Anshuman Maharana, Satchitananda Naik, Arkady A. Tseytlin, Andrej Stepanchuk and Soumya Bhattacharya for their comments. The authors gratefully acknowledge the kind hospitality of CHEP, IISc Bangalore and ICTS-TIFR during Advanced String School 2015 and Strings conference 2015, where a major part of this work was done.

 \bibliographystyle{JHEP}
 \bibliography{ads3s3s3}

\end{document}